\documentclass[11pt]{article}
\usepackage{graphics,epsfig, psfrag}
\usepackage{amsfonts,amssymb}

\begin{document}

\setcounter{page}{1}
\vfil

\pagestyle{plain}

\begin{center}
{\LARGE {\bf A New Quantization Map}}

\bigskip
\bigskip
\bigskip

{\large D. Mauro}

             Dipartimento di Fisica Teorica, 
	     Universit\`a di Trieste, \\
	     Strada Costiera 11, P.O.Box 586, Trieste, Italy,\\
	     and INFN, Sezione di Trieste, Italy\\
	     e-mail: {\it mauro@ts.infn.it}

\end{center}

\bigskip
\bigskip
\begin{abstract}
\noindent In this paper we find a simple rule to reproduce the algebra of 
quantum observables using only the commutators and operators which appear 
in the Koopman-von Neumann (KvN) formulation of classical mechanics.
The usual Hilbert space of quantum mechanics becomes embedded in the KvN 
Hilbert space: in particular it turns out to be the subspace on which the 
quantum positions Q and momenta P act irreducibly. 
\end{abstract}

\section{Introduction}

It is well known that in classical statistical mechanics the evolution of the probability
densities in phase space $\rho(q,p,t)$ is given by the Liouville equation
\begin{equation}
\displaystyle i\frac{\partial}{\partial t}\rho(q,p,t)=\hat{L}\rho(q,p,t) \label{Liorho}
\end{equation}
where $\hat{L}$ is the Liouville operator
\begin{equation}
\hat{L}= -i\partial_pH(q,p)\partial_q+i\partial_qH(q,p)\partial_p \label{liouv}
\end{equation}
and $H(q,p)$ is the Hamiltonian in the phase space ${\cal M}$ of the system. In 
\cite{Koopman} and \cite{von Neumann} KvN formulated classical mechanics 
in a Hilbert space made up
of complex square integrable functions over the phase space variables $\psi(q,p,t)$.  
In particular they postulated, as
equation of evolution for $\psi(q,p,t)$, the Liouville equation itself:
\begin{equation}
\displaystyle 
i\frac{\partial}{\partial t}\psi(q,p,t)=\hat{L}\psi(q,p,t). \label{Liopsi}
\end{equation}
Starting from (\ref{Liopsi}) it is 
easy to prove that, since the Liouvillian $\hat{L}$ contains only first order derivatives,
the Liouville equation (\ref{Liorho}) for the probability densities
$\rho(q,p,t)$ can be derived via the postulate $\rho(q,p,t)=|\psi(q,p,t)|^2$.
Finally KvN imposed on the states of their Hilbert space the following scalar
product:
\begin{equation}
\langle \psi|\tau\rangle=\int dqdp \,\psi^*(q,p)\tau(q,p). 
\label{scalar}
\end{equation}
With this choice the Liouvillian $\hat{L}$ is a Hermitian
operator. Therefore $\langle\psi|\psi\rangle=\int dqdp|\psi(q,p)|^2$ is a conserved
quantity and $|\psi(q,p)|^2$ can be consistently interpreted as the probability 
density of finding a particle in a point of the phase space. 

Now every theory formulated via states and operators in a Hilbert space can be formulated 
also via path integrals. It was proved in Ref. \cite{gozzi} and   
\cite{waves1} that it is possible to describe the evolution (\ref{Liopsi}) in the following 
manner:
$ \displaystyle \psi(\varphi,t)=\int d\varphi_i\, 
\langle\varphi,t|\varphi_i,t_i\rangle\psi(\varphi_i,t_i)$
where we have indicated with $\varphi\equiv (q,p)$ all the phase space variables. 
The kernel of propagation $\langle\varphi,t|\varphi_i,t_i\rangle$ 
has the following path integral expression:
\begin{equation}
\displaystyle \langle\varphi,t|\varphi_i,t_i\rangle=\int {\cal D}^{\prime\prime}\varphi
{\cal D}\lambda \,\textrm{exp}\biggl[i\int dt\, (\lambda_a\dot{\varphi}^a-L)
\biggr] \label{path}
\end{equation}
where the double prime in ${\cal D}^{\prime\prime}\varphi$ 
indicates that the integration is over paths with fixed end points in $\varphi$. 
The $L$ in the weight of the path integral (\ref{path}) is given by:
\begin{equation}
L=\lambda_a\omega^{ab}\partial_bH\equiv \lambda_q\partial_pH-\lambda_p\partial_qH 
\label{liouvilliano}.
\end{equation}
Having a path integral we can introduce the concept of a commutator as Feynman
did in the quantum case: given two functions $O_{\scriptscriptstyle 1}(\varphi,\lambda)$ and
$O_{\scriptscriptstyle 2}(\varphi,\lambda)$, we can evaluate the following quantity under the path integral
(\ref{path}): $\displaystyle [\hat{O}_{\scriptscriptstyle 1},\hat{O}_{\scriptscriptstyle 2}]
\equiv\lim_{\epsilon\to 0}\langle
O_{\scriptscriptstyle 1}(t+\epsilon)O_{\scriptscriptstyle 2}(t)-O_{\scriptscriptstyle 2}
(t+\epsilon)O_{\scriptscriptstyle 1}(t)\rangle$. 
What we get is \cite{gozzi}:
\begin{equation}
[\hat{\varphi}^a,\hat{\varphi}^b]=0,\qquad\quad [\hat{\lambda}_a,\hat{\lambda}_b]=0,
\qquad\quad [\hat{\varphi}^a,\hat{\lambda}_b]=i\delta_b^a. \label{comm}
\end{equation}
The first commutator of (\ref{comm}) tells us that the positions $\hat{q}$ commute with the momenta $\hat{p}$, i.e.
that we are doing classical and not quantum mechanics. 
The last commutator instead tells us the $\hat{\lambda}_a$ are something like
the momenta conjugate to $\hat{\varphi}^a$. In order to satisfy (\ref{comm}) 
we can use the representation in which $\hat{\varphi}^a$ is a multiplicative operator and 
$\hat{\lambda}_a$ a derivative one:
\begin{equation}
\displaystyle \hat{\varphi}^a=\varphi^a,\qquad\qquad\qquad 
\hat{\lambda}_a=-i\frac{\partial}{\partial \varphi^a}. \label{realiz}
\end{equation}
Via the previous operatorial realization, the $L$ of (\ref{liouvilliano}) can also be turned
into an operator:
$L\,\rightarrow\, \hat{L}=-i\omega^{ab}\partial_bH\partial_a=
-i\partial_pH\partial_q+i\partial_qH\partial_p.$
Therefore $\hat{L}$ is just the Liouville operator of Eq. (\ref{liouv}). 
This confirms that the operatorial
formalism lying behind the path integral (\ref{path}) is nothing more than the KvN one.
 
The main problem we are interested in this paper is the
quantization of classical mechanics, once it is formulated 
via operatorial or path integral techniques. We know that usually the quantization of a system is
performed via the Dirac's correspondence rules, i.e. by replacing the
classical
Poisson brackets $\{\,\cdot\,,\,\cdot\,\}_{Pb}$ by commutators according to the
following relation:
\begin{displaymath}
\displaystyle \{\,\cdot\,,\,\cdot\,\}_{Pb}\;\longrightarrow\;
\frac{[\,\cdot\,,\,\cdot\,]_{\scriptscriptstyle QM}}{i\hbar}.
\end{displaymath}
Now, in classical mechanics formulated \`a la KvN, the Poisson brackets are already replaced by
the KvN commutators (\ref{comm}).
So the quantization of the system can be performed:\newline 
{\bf 1)} either finding suitable
rules to go from 
the KvN commutators to the quantum ones;\newline 
{\bf 2)} or finding a way to reproduce the
algebra of quantum
observables using KvN commutators and operators.\newline
In this paper we will find a compact way to implement the second alternative,
by using the fact that the set of Hermitian operators in the KvN theory 
includes also operators depending on $\hat{\lambda}$ and not only the functions $f(\hat{q},\hat{p})$. 
In particular in {\bf Sec. 2} we will introduce the following map:
\begin{equation}
\displaystyle {\mathcal Q}:\;f(\hat{q},\hat{p})\;\longrightarrow\;
f\Bigl(\hat{q}-\frac{\hbar}{2}\hat{\lambda}_p,\hat{p}+\frac{\hbar}{2}\hat{\lambda}_q\Bigr)
\label{mapq}
\end{equation}
where the domain of ${\mathcal Q}$ is made up of all the standard classical
observables
$f(\hat{q},\hat{p})$ while the image of ${\mathcal Q}$ is made up of suitable
Hermitian operators living in the KvN Hilbert space and depending also on
$\hat{\lambda}$.
Using the KvN commutators (\ref{comm}) all the functions of the form $\displaystyle
f\Bigl(\hat{q}-\frac{\hbar}{2}\hat{\lambda}_p,\hat{p}+\frac{\hbar}
{2}\hat{\lambda}_q\Bigr)$ which appear on the RHS of (\ref{mapq}) 
satisfy exactly the algebra of the 
observables of quantum mechanics. So we can say that, among all the operators of the KvN 
Hilbert space, there is a set whose algebra is isomorphic to the algebra of quantum observables.
The most interesting feature of our approach is the split between 
the {\it quantum} observables of the RHS of (\ref{mapq}) and the {\it classical} ones $f(\hat{q},\hat{p})$.
In particular in {\bf Sec. 3} we will show that, while the classical energy $H(\hat{q},\hat{p})$ 
commutes with the Liouvillian $\hat{L}$, the associated quantum energy 
is not conserved under the evolution generated by $\hat{L}$.
Therefore, in order to preserve the conservation of energy, we must replace also the Liouvillian 
$\hat{L}$ with a more complicated object $\hat{\cal G}$ which is well known in the literature 
because it is the Moyal operator which appears in the equation of evolution of the Wigner functions
\cite{quantoptics}. 
This replacement is the first step necessary in order to go from the classical path integral 
(\ref{path}) to the quantum one. The second step is a sort of polarization prescription 
which reduces the KvN Hilbert space to the standard Hilbert space of quantum mechanics. 
In {\bf Sec. 4} we will see how this polarization procedure can be justified 
by requiring the irreducibility of the Hilbert space of quantum mechanics under the 
action of the {\it quantum} observables. In the {\bf Conclusions}
we will outline some of the possible future applications of the content of this paper.

\section{From Classical to Quantum Observables}

Usually physicists identify the observables of classical mechanics 
with the functions of the phase space variables $f(\varphi)$. In the operatorial
approach to classical mechanics such observables become the
functions of the commuting operators $\hat{\varphi}^a$. Therefore 
the algebra of classical observables is Abelian and
there is no uncertainty principle involving the operators
$\hat{\varphi}^a$ which can be diagonalized 
simultaneously. Every simultaneous eigenstate of $\hat{\varphi}^a$ 
determines uniquely a point in the phase space $|\varphi^a_{\scriptscriptstyle (0)}\rangle$
which represents the state of the system. On such a state all the observables of classical
mechanics assume a well defined value. In fact if
$\hat{\varphi}|\varphi^a_{\scriptscriptstyle (0)}\rangle=
\varphi^a_{\scriptscriptstyle (0)}|\varphi^a_{\scriptscriptstyle (0)}\rangle$
then $f(\hat{\varphi})|\varphi^a_{\scriptscriptstyle (0)}\rangle=
f(\varphi^a_{\scriptscriptstyle (0)})|\varphi^a_{\scriptscriptstyle
(0)}\rangle$,
i.e. the eigenstate $|\varphi^a_{\scriptscriptstyle (0)}\rangle$ of
$\hat{\varphi}$ is an eigenstate also for all the observables $f(\hat{\varphi})$
and the associated eigenvalue is just the function $f$ evaluated on the point
of the phase space $\varphi^a_{\scriptscriptstyle (0)}$. In this approach the mean values
and the probability distributions of the classical observables are completely independent of the phases
of the wave functions $\psi(\varphi)$ and only the modulus $|\psi(\varphi)|$ is significant 
from a physical point of view \cite{kvn2}.

Besides the operators $\hat{\varphi}^a$ in the operatorial formulation of classical mechanics 
we have also the operators $\hat{\lambda}_a$ which allow us to generate a non trivial
evolution via the Liouvillian 
$L$ of Eq. (\ref{liouvilliano}). Such operators 
$\hat{\lambda}_a$ do not commute with $\hat{\varphi}^a$ and this immediately
suggests that it might be possible to construct a non Abelian algebra of operators
by considering suitable combinations of 
$\hat{\varphi}$ and $\hat{\lambda}$. In particular it would be very
interesting to discover whether it is possible to reproduce, using only the
KvN commutators (\ref{comm}), the non Abelian algebra of quantum observables.
In order to do this we can define the following operators, known in the literature 
on Wigner functions as Bopp operators, \cite{quantoptics}:
\begin{equation}
\left\{
\begin{array}{l}
\displaystyle \hat{Q}_j\equiv \hat{q}_j-\frac{1}{2}\hbar\hat{\lambda}_{p_j}\medskip\\
\displaystyle \hat{P}_j\equiv \hat{p}_j+\frac{1}{2}\hbar\hat{\lambda}_{q_j}.
\end{array}
\right.
\label{rules}
\end{equation}
Via the KvN commutators (\ref{comm}) it is very easy to prove that they 
satisfy the usual Heisenberg algebra: $[\hat{Q}_j,\hat{P}_k]=i\hbar\delta_{jk}$. 
Since both $\hat{\varphi}$ and $\hat{\lambda}$ are Hermitian under the KvN scalar
product (\ref{scalar}) 
we have that also $\hat{Q}_j$ and $\hat{P}_j$ 
are Hermitian operators with the same scalar product. Not only, but the {\it classical} operators 
$\hat{q}_j$ and $\hat{p}_j$ can be obtained from the associated {\it quantum} ones
in the limit $\hbar\to 0$. We would like to stress that the {\it quantum} positions and momenta are now different 
``entities" than the {\it classical} ones. This may be difficult 
to accept but at the same time it may be the crucial element which makes quantum mechanics 
so counter-intuitive and different from classical mechanics.

The rules (\ref{rules}) can be easily generalized
and the quantum operator $\hat{F}$ associated to the classical observable
$f(\hat{q},\hat{p})$ will be given by the following function:
\begin{equation}
\displaystyle \hat{F}=f\biggl(\hat{q}_j-\frac{1}{2}\hbar\hat{\lambda}_{p_j},
\hat{p}_j+\frac{1}{2}\hbar\hat{\lambda}_{q_j}\biggr). \label{capeffe}
\end{equation}
Since $\hat{\varphi}$ and $\hat{\lambda}$ do not commute it is clear that in (\ref{capeffe}) we must specify 
also the ordering used. The one we choose, once we expand (\ref{capeffe}) in $\hbar$, is the following:
\begin{equation}
\displaystyle \hat{F}=f(\hat{q},\hat{p})+
\sum_{n=1}^{\infty}\frac{1}{n!}\biggl(\frac{\hbar}{2}\biggr)^n\hat{\lambda}_{a_1}
\cdots\hat{\lambda}_{a_n}\omega^{a_1b_1}\cdots \omega^{a_nb_n}\partial_{b_1}
\cdots \partial_{b_n}f. \label{expansion}
\end{equation}
The operators $\hat{F}$ are Hermitian under the KvN scalar product (\ref{scalar}) and,
since the basic commutators
$[\hat{Q}_j,\hat{P}_k]=i\hbar\delta_{jk}$ are satisfied by the
choice (\ref{rules}), the Hermitian functions of
$\hat{Q}_j$ and $\hat{P}_j$ given by (\ref{capeffe}) or (\ref{expansion}) will satisfy 
the correct quantum commutators. 
For example the three components of the angular momentum turn out to be:
\begin{displaymath}
\begin{array}{l}
\displaystyle \hat{M}_x=\hat{y}\hat{p}_z-\hat{z}\hat{p}_y-\frac{1}{2}\hbar\bigl[\hat{\lambda}_y\hat{z}-\hat{\lambda}_{z}
\hat{y}+\hat{\lambda}_{p_y}\hat{p}_z-\hat{\lambda}_{p_z}\hat{p}_y\bigr]
-\frac{1}{4}\hbar^2\bigl[\hat{\lambda}_{p_y}\hat{\lambda}_z-\hat{\lambda}_y\hat{\lambda}_{p_z}\bigr]\medskip \\ 
\displaystyle \hat{M}_y=\hat{z}\hat{p}_x-\hat{x}\hat{p}_z-\frac{1}{2}\hbar\bigl[\hat{\lambda}_z\hat{x}-\hat{\lambda}_{x}
\hat{z}+\hat{\lambda}_{p_z}\hat{p}_x-\hat{\lambda}_{p_x}\hat{p}_z\bigr]
-\frac{1}{4}\hbar^2\bigl[\hat{\lambda}_{p_z}\hat{\lambda}_x-\hat{\lambda}_z\hat{\lambda}_{p_x}\bigr]\medskip \\ 
\displaystyle \hat{M}_z=\hat{x}\hat{p}_y-\hat{y}\hat{p}_x-\frac{1}{2}\hbar\bigl[\hat{\lambda}_x\hat{y}-\hat{\lambda}_{y}
\hat{x}+\hat{\lambda}_{p_x}\hat{p}_y-\hat{\lambda}_{p_y}\hat{p}_x\bigr]
-\frac{1}{4}\hbar^2\bigl[\hat{\lambda}_{p_x}\hat{\lambda}_y-\hat{\lambda}_x\hat{\lambda}_{p_y}\bigr]
\end{array}
\end{displaymath}
and they obey the usual algebra $[\hat{M}_i,\hat{M}_j]=i\hbar \epsilon_{ijk}\hat{M}_k$ under the KvN commutators
(\ref{comm}).

All this confirms that it is possible to build the quantum observables
$\hat{F}$ by adding to the associated classical ones $f(\hat{q},\hat{p})$ a suitable 
combination of KvN operators which vanishes in the classical limit:
$\displaystyle \lim_{\hbar\to 0}\hat{F}=f(\hat{q},\hat{p})$.

\section{From Classical to Quantum Evolutions}

As we have seen in the previous section, in classical mechanics
the observable energy is given by the operator $H(\hat{q},\hat{p})$.
According to the map ${\mathcal Q}$ of (\ref{mapq}) 
at the quantum level the energy has instead to be identified
with the operator $H(\hat{Q},\hat{P})$ where the classical position $\hat{q}$
and momentum 
$\hat{p}$ are replaced by the associated Bopp operators $\hat{Q}$ 
and $\hat{P}$. Now at the classical level the energy $H(\hat{q},\hat{p})$ is a conserved
quantity since it commutes with the Liouvillian $\hat{L}$ which is the generator of the 
time evolution: $[\hat{L},H(\hat{q},\hat{p})]=-i\partial_aH\omega^{ab}
\partial_bH=0.$
What happens at the quantum level? Does the operator $H(\hat{Q},\hat{P})$ 
still commute with the Liouvillian $\hat{L}$? The answer is no. In fact, using (\ref{expansion}),
we have that the commutator of the Liouvillian with the quantum energy $H(\hat{Q},\hat{P})$ is:
\begin{eqnarray}
 \displaystyle \bigl[\hat{L},H(\hat{Q},\hat{P})\bigr]&=&-\frac{\hbar^2}{8}\omega^{\alpha\beta}
\omega^{a_{\scriptscriptstyle 1}b_{\scriptscriptstyle 1}}\omega^{a_{\scriptscriptstyle 2}
b_{\scriptscriptstyle 2}}(i\hat{\lambda}_{a_{\scriptscriptstyle 1}}
\hat{\lambda}_{a_{\scriptscriptstyle 2}}
\partial_{\alpha}\partial_{b_{\scriptscriptstyle 1}}
\partial_{b_{\scriptscriptstyle 2}}H\partial_{\beta}H+\nonumber\\
&&+\hat{\lambda}_{\alpha}\partial_{\beta}
\partial_{a_{\scriptscriptstyle 1}}
\partial_{a_{\scriptscriptstyle 2}}H\partial_{b_{\scriptscriptstyle 1}}
\partial_{b_{\scriptscriptstyle 2}}H)+O(\hbar^3). \label{notcons}
\end{eqnarray}
For Hamiltonians more than quadratic in $q$ and $p$ the previous commutator is different from
zero and the energy is not conserved under the evolution generated by the Liouvillian. 
Therefore if we identify the quantum energy with $H(\hat{Q},\hat{P})$ and we require that this 
quantity has to be conserved in time, then
we must consistently change also the operator which generates the time evolution. 
In particular in order to compensate the RHS of (\ref{notcons}) we must add to the Liouvillian
a term cubic in $\lambda$ and in the symplectic matrix: $\hat{\cal G}_{\scriptscriptstyle 
(1)}=\displaystyle \frac{1}{24}
\hbar^2 \hat{\lambda}_a\hat{\lambda}_b\hat{\lambda}_c
\omega^{ad}\omega^{be}\omega^{cf}\partial_d\partial_e\partial_fH$.
In fact the commutator of such a term with $H(\hat{Q},\hat{P})$:
\begin{eqnarray}
\displaystyle \bigl[\hat{\cal G}_{\scriptscriptstyle 
(1)},H(\hat{Q},\hat{P})\bigr]&=& 
\displaystyle \frac{\hbar^2}{8}\omega^{\alpha\beta}
\omega^{a_{\scriptscriptstyle 1}b_{\scriptscriptstyle 1}}
\omega^{a_{\scriptscriptstyle 2}b_{\scriptscriptstyle 2}}
(i\hat{\lambda}_{a_{\scriptscriptstyle 1}}\hat{\lambda}_{a_{\scriptscriptstyle 2}}
\partial_{\alpha}\partial_{b_{\scriptscriptstyle 1}}
\partial_{b_{\scriptscriptstyle 2}}H\partial_{\beta}H+\nonumber\\
&&+\hat{\lambda}_{\alpha}\partial_{\beta}
\partial_{a_{\scriptscriptstyle 1}}
\partial_{a_{\scriptscriptstyle 2}}H\partial_{b_{\scriptscriptstyle 1}}
\partial_{b_{\scriptscriptstyle 2}}H)+O(\hbar^3) \nonumber
\end{eqnarray}
is just the opposite of the RHS of (\ref{notcons}). If we consider also the other
terms of the expansion in $\hbar$ we have that $H(\hat{Q},\hat{P})$ is conserved if
we replace the Liouvillian $\hat{\lambda}_a\omega^{ab}\partial_bH$ with the following operator:
\begin{equation}
\label{arr}
\displaystyle \hat{\cal G}=
\sum_{j=0}^{\infty}\frac{\hbar^{2j}}{2^{2j}(2j+1)!}\hat{\lambda}_{a_{\scriptscriptstyle 1}}\cdots
\hat{\lambda}_{a_{\scriptscriptstyle 2j+1}} 
\omega^{a_{\scriptscriptstyle 1}b_{\scriptscriptstyle 1}}\cdots\omega^{a_{\scriptscriptstyle 2j+1}b_{
\scriptscriptstyle 2j+1}}\partial_{b_{\scriptscriptstyle 1}}
\cdots\partial_{b_{\scriptscriptstyle 2j+1}}H.
\end{equation}
From (\ref{arr}) we see that $\hat{\cal G}$ is given by the Liouvillian ($j=0$)
plus a sum of terms which are 
proportional to the even powers of $\hbar$ and in the limit $\hbar\to 0$ we have
that $\hat{\cal G}$ reduces to the Liouvillian $\hat{L}$ itself. 
At the path integral level this first step of replacing $\hat{L}$ 
with $\hat{\cal G}$ implies that the kernel of propagation (\ref{path}) must be replaced by
\cite{marinov}:
\begin{equation}
\langle \varphi,t|\varphi_i,t_i\rangle=\int {\cal D}^{\prime\prime}\varphi
{\cal D}\lambda\;\textrm{exp}\biggl[i\int dt (\lambda_a\dot{\varphi}^a-{\cal G})\biggr].
\label{kernelg}
\end{equation}
If we include, besides $\lambda$ and $\varphi$, also the forms $d\varphi$ and the 
vector fields then the associated path
integral has been worked out in Ref. \cite{ennio}.
Since the kinetic term of (\ref{kernelg}) is the usual one, $\lambda_a\dot{\varphi}^a$,
the KvN commutators 
(\ref{comm}) and their operatorial realization
(\ref{realiz}) are unchanged. 
If we limit ourselves to the case of a 2D phase space with a Hamiltonian quadratic in the momenta,
$\displaystyle H(q,p)=\frac{p^2}{2m}+V(q)$,
then the operatorial realization of ${\cal G}$ becomes:
\begin{eqnarray}
\displaystyle \hat{\cal G}&=&\hat{L}-\sum_{j=1}^{\infty}\biggl(\frac{\hbar}{2}\biggr)^{2j}
\frac{1}{(2j+1)!}(\hat{\lambda}_{p})^{2j+1}\frac{d^{2j+1}}{dq^{2j+1}}V(q)=\nonumber \\
\displaystyle &=&-i\frac{p}{m}\frac{\partial}{\partial q}+i
\sum_{j=0}^{\infty}\biggl(\frac{\hbar}{2}\biggr)^{2j}\frac{(-1)^j}{(2j+1)!}
\frac{d^{2j+1}}{d q^{2j+1}}V(q)
\frac{\partial^{2j+1}}{\partial p^{2j+1}}.\nonumber
\end{eqnarray}
Before going on it is interesting to note that $\hat{\cal G}$ is just the operator which appears 
in the equation of evolution of the Wigner functions $W(q,p,t)$, see Ref. \cite{quantoptics}:
$\displaystyle i\frac{\partial}{\partial t}W(q,p,t)=\hat{\cal G}W(q,p,t)$.
In any case, in order to avoid possible misunderstandings, we want to stress that in our approach 
the operator $\hat{\cal G}$ replaces the Liouvillian $\hat{L}$ in making the evolution of the
elements of the KvN Hilbert space, which are the probability amplitudes $\psi(q,p,t)$  
and not the quasi-probability distributions $W(q,p,t)$ of the Wigner approach.  

Let us now define, besides the Bopp operators of Eq. (\ref{rules}), the following operators:
\begin{equation}
\left\{
\begin{array}{l}
\displaystyle \hat{\bar{Q}}_j\equiv\hat{q}_j+\frac{1}{2}\hbar\hat{\lambda}_{p_j}\medskip\\
\displaystyle \hat{\bar{P}}_j\equiv\hat{p}_j-\frac{1}{2}\hbar \hat{\lambda}_{q_j}.
\end{array}
\right. \label{delta}
\end{equation}
It is then possible to prove \cite{Kubo} that the $\hat{\cal G}$ of (\ref{arr}) can be written in terms of the standard 
Hamiltonian $H$ as:
\begin{equation}
\displaystyle \hat{\cal G}=\frac{1}{\hbar}H(\hat{Q},\hat{P})-\frac{1}{\hbar}
H(\hat{\bar{Q}},\hat{\bar{P}}) \label{difference}
\end{equation}
where the expansion in $\hbar$ of $H(\hat{Q},\hat{P})$ is given by the RHS of (\ref{expansion})
with $f$ replaced by $H$ while the expansion in $\hbar$ of $H(\hat{\bar{Q}},\hat{\bar{P}})$
is:
\begin{displaymath}
H(\hat{\bar{Q}},\hat{\bar{P}})=H(\hat{q},\hat{p})+
\sum_{n=1}^{\infty}\frac{1}{n!}
\biggl(-\frac{\hbar}{2}\biggr)^n\hat{\lambda}_{a_1}
\cdots \hat{\lambda}_{a_n}\omega^{a_1b_1}\cdots \omega^{a_nb_n}\partial_{b_1}
\cdots \partial_{b_n}H. 
\end{displaymath}
From (\ref{difference}) it is also easy to understand
the reason why the energy $H(\hat{Q},\hat{P})$ commutes with the operator $\hat{\cal{G}}$.
In fact $\hat{\cal G}$ is proportional to the difference of the quantum energy itself $H(\hat{Q},\hat{P})$
and an operator, $H(\hat{\bar{Q}},\hat{\bar{P}})$, which commutes with the quantum energy 
because the unbarred operators $(\hat{Q},\hat{P})$ of (\ref{rules})
always commute with the barred ones $(\hat{\bar{Q}},\hat{\bar{P}})$ of (\ref{delta}).

Since the operator of evolution $\hat{\cal G}$ is now expressed via
the operators (\ref{rules})-(\ref{delta}), it is quite natural to write down 
also the path integral (\ref{kernelg}) in terms of $Q,\bar{Q},P$ and $\bar{P}$. 
Since $\hat{q}$ and $\hat{\lambda}_p$ 
are coupled together in the definition of $\hat{Q}$ and $\hat{\bar{Q}}$,
it is more convenient to use the representation in which the 
multiplicative operators are
$\hat{q}$ and $\hat{\lambda}_p$ instead of $\hat{q}$ and $\hat{p}$
\cite{ann}. In this representation the KvN Hilbert space is made up of 
functions of $(q,\lambda_p)$ and their kernel of propagation becomes:
\begin{eqnarray}
\label{beta}
\displaystyle \langle q,\lambda_p,t|q_i,\lambda_{p_i},t_i\rangle &=& 
\int \frac{dp}{\sqrt{2\pi}}\frac{dp_i}{\sqrt{2\pi}} 
\,e^{-i\lambda_pp}\langle\varphi,t|\varphi_i,t_i\rangle e^{i\lambda_{p_i}p_i}=\\
&=&\int {\cal D}^{\prime\prime}q{\cal D}^{\prime\prime}\lambda_p{\cal D} p
{\cal D}\lambda_q\;\textrm{exp}\biggl[i\int dt
(\lambda_q\dot{q}-p\dot{\lambda}_p-{\cal G})\biggr]. \nonumber
\end{eqnarray}
The inverse of (\ref{rules})-(\ref{delta}) are given by:
\begin{displaymath}
\left\{
\begin{array}{l}
\displaystyle \hat{q}=\frac{\hat{Q}+\hat{\bar{Q}}}{2}\medskip\\
\displaystyle \hat{\lambda}_p=\frac{\hat{\bar{Q}}-\hat{Q}}{\hbar} 
\end{array}
\right. \qquad\qquad
\left\{
\begin{array}{l}
\displaystyle \hat{p}=\frac{\hat{P}+\hat{\bar{P}}}{2}\medskip\\
\displaystyle \hat{\lambda}_q=\frac{\hat{P}-\hat{\bar{P}}}{\hbar}.
\end{array}
\right. 
\end{displaymath}
Using them we can rewrite the kernel of propagation (\ref{beta}) as:
\begin{eqnarray}
\langle Q,\bar{Q},t|Q_i,\bar{Q}_i,t_i\rangle &\hspace{-0.2cm}=&\hspace{-0.2cm}
\int {\cal D}^{\prime\prime}Q\frac{{\cal D}P}{\hbar}
\; \textrm{exp}\,\biggl[\frac{i}{\hbar}\int dt \bigl[P\dot{Q}-H(Q,P)\bigr]\biggr]\cdot\nonumber\\
&\hspace{-0.2cm}&\hspace{-0.2cm} \int {\cal D}^{\prime\prime}\bar{Q}\frac{{\cal D}\bar{P}}{\hbar}\;
\textrm{exp}\, \biggl[-\frac{i}{\hbar}\int dt 
\bigl[\bar{P}\dot{\bar{Q}}-H(\bar{Q},\bar{P})\bigr]\biggr]. \label{trentuno}
\end{eqnarray}
This path integral generates the evolution of the square integrable 
functions $\psi(Q,\bar{Q})$. From (\ref{trentuno}) it is easy to note that the evolution of the variables $(Q,P)$  
is completely decoupled from the evolution of the variables $(\bar{Q},\bar{P})$. 
So if we consider as initial wave functions those $\psi(Q,\bar{Q})$ which can be factorized as 
$\psi(Q,\bar{Q})=\bar{\psi}(Q)\cdot \widetilde{\psi}(\bar{Q})$, then they will remain 
factorized during the time evolution. In fact the variables $(Q,P)$ and $(\bar{Q},\bar{P})$ 
cannot become entangled because in (\ref{trentuno}) there is no interaction between them. 
Furthermore if we limit the KvN Hilbert space to the one 
whose basis is given by all the eigenstates $|Q\rangle$ of the operator $\hat{Q}$, then from (\ref{trentuno}) 
we have that the wave functions $\bar{\psi}(Q)$ evolve just with the usual quantum kernel of
evolution for a system described by a Hamiltonian $H(Q,P)$:
\begin{displaymath}
\langle Q,t|Q_i,t_i\rangle=\int {\cal D}^{\prime\prime}Q\frac{{\cal D}P}{\hbar}\, \textrm{exp}\,
\biggl[\frac{i}{\hbar}
\int dt [P\dot{Q}-H(Q,P)]\biggr].
\end{displaymath}
In the next section we will analyze the reasons why we have to 
restrict the KvN Hilbert space to the usual one in the quantum case
and we will show that also this prescription is just a consequence 
of the choice of the observables (\ref{mapq}).

%

\section{From Classical to Quantum States}

In this section we will show how the usual Hilbert space of quantum
mechanics can be embedded in the KvN Hilbert space. Let us suppose 
we use in quantum mechanics the coordinate representation:
the positions $\hat{Q}$ are operators 
of multiplication by $Q$, the momenta $\hat{P}$ are given by 
$\displaystyle \hat{P}=-i\hbar\frac{\partial}{\partial Q}$, 
while the Hilbert space is made up of the complex and square integrable 
functions of $Q$. How can we obtain all this starting from the KvN states and operators?
As we have already seen in the previous sections, 
since $\hat{Q}$ is a linear combination of $\hat{q}$ and $\hat{\lambda}_{p}$,
it is more convenient to use, at the KvN Hilbert space level, the representation 
in which $\hat{q}$ and $\hat{\lambda}_{p}$ are multiplicative 
operators while, to satisfy  (\ref{comm}),
$\hat{\lambda}_q$ and $\hat{p}$ must be given by
$\displaystyle \hat{\lambda}_{q}\equiv -i\frac{\partial}{\partial q}$
and $\displaystyle \hat{p}\equiv i\frac{\partial}{\partial\lambda_{p}}$
respectively. In this representation the Hilbert space of the theory 
is made up of the complex wave functions 
$\psi(q,\lambda_p)$ which are square integrable under the KvN scalar product  
$\langle \psi|\tau\rangle=\int dq d\lambda_p 
\psi^*(q,\lambda_p)\tau(q,\lambda_p)$.
Equivalently, if we use the variables $Q$ and $\bar{Q}$, the KvN Hilbert space will be given by
all the wave functions
$\psi(Q,\bar{Q})$ which are square integrable under the scalar product:
$ \langle \psi|\tau\rangle=\int dQd\bar{Q}\,\psi^*(Q,\bar{Q})
\tau(Q,\bar{Q})$. The quantum position  
$\displaystyle \hat{Q}=\hat{q}-\frac{1}{2}\hbar\hat{\lambda}_{p}$ multiplies a wave function
$\psi(Q,\bar{Q})$  by $\displaystyle Q=q-\frac{1}{2}\hbar\lambda_p$
while $\displaystyle \hat{P}=\hat{p}+\frac{1}{2}\hbar\hat{\lambda}_q$ becomes the following 
operator: $\displaystyle \hat{P}=i\frac{\partial}{\partial \lambda_{p}}-\frac{i}{2}
\hbar\frac{\partial}{\partial q}$. It is easy to realize that $\hat{Q}$
and $\hat{P}$ act on the part in $Q$ of the wave functions $\psi(Q,\bar{Q})$ just as the quantum 
observables act on the standard wave functions of quantum mechanics. 
In fact:
\begin{displaymath}
\left\{
\begin{array}{l}
\hat{Q}\psi(Q,\bar{Q})= Q\psi(Q,\bar{Q}) \medskip \\ 
\displaystyle \hat{P}\psi(Q,\bar{Q})= 
\biggl(i\frac{\partial}{\partial\lambda_p}-
\frac{i}{2}\hbar\frac{\partial}{\partial q}\biggr)\psi(Q,\bar{Q})=
-i\hbar\frac{\partial}{\partial Q}\psi(Q,\bar{Q}). 
\end{array}
\right.
\end{displaymath}
In general the action of the quantum observable $\hat{F}$ on $\psi(Q,\bar{Q})$ is given by:
\begin{equation}
\displaystyle \hat{F}\psi(Q,\bar{Q})=f(\hat{Q},\hat{P})\psi(Q,\bar{Q})=
f\biggl(Q,-i\hbar\frac{\partial}{\partial Q}\biggr)
\psi(Q,\bar{Q}). \label{effe}
\end{equation}
Clearly the space of the complex and square integrable wave functions $\psi(Q,\bar{Q})$ 
is too big in order to be an irreducible representation of the algebra of quantum 
observables. In fact, from (\ref{effe}) we see that every quantum observable $\displaystyle 
\hat{F}$ modifies only the part in $Q$ 
of the wave function $\psi(Q,\bar{Q})$ while it acts as the identity operator on the part in $\bar{Q}$. Therefore
if we require 
that the quantum operators $\hat{Q}$ and $\hat{P}$ 
act {\it irreducibly} (see page 434 of \cite{Abraham}) on a space of states 
${\bf H}$ then we cannot identify ${\bf H}$ with the full
KvN Hilbert space of the square integrable wave functions $\psi(Q,\bar{Q})$. In fact there exists at least 
one non-trivial subspace 
of the KvN Hilbert space which is invariant under the action of the quantum positions
$\hat{Q}$, the quantum momenta $\hat{P}$ and of every other quantum observable
$\displaystyle \hat{F}=f\biggl(Q,-i\hbar\frac{\partial}{\partial Q}\biggr)$. 
This non-trivial subspace is given by the set of {\it all} the possible square integrable 
functions in $Q$ times one {\it particular} square integrable function in $\bar{Q}$,
let us call it $\chi(\bar{Q})$:
\begin{equation}
\displaystyle 
{\bf H}_{\chi}=\biggl\{ \psi(Q)\chi(\bar{Q})\; \textrm{with} \, \int dQd\bar{Q}\,
|\psi(Q)|^2|\chi(\bar{Q})|^2=1\biggr\}. \label{aro}
\end{equation}
The Hermitian quantum observables $\displaystyle f\biggl(Q,-i\hbar\frac{\partial}{\partial Q}\biggr)$ 
do not modify the function $\chi(\bar{Q})$ while they map a 
wave function of the form $\psi(Q)$ in another of the same type $\psi^{\prime}(Q)$. 
In other words the quantum 
observables map every vector of ${\bf H}_{\chi}$ in another vector of ${\bf H}_{\chi}$. Since the
function $\chi(\bar{Q})$ is fixed the space ${\bf H}_{\chi}$ is isomorphic 
to the usual Hilbert space of quantum mechanics:
\begin{equation}
{\bf H}=\biggl\{\psi(Q)\; \textrm{with} \int dQ |\psi(Q)|^2=1\biggr\}. \label{quantumhil}
\end{equation} 
Also the KvN scalar product between two different states of ${\bf H}_{\chi}$, let us say $\psi(Q)\chi(\bar{Q})$ and
$\tau(Q)\chi(\bar{Q})$:
\begin{displaymath}
\langle \psi|\tau\rangle=\int dQd\bar{Q}\;\psi^*(Q)\chi^*(\bar{Q})\tau(Q)\chi(\bar{Q})
\end{displaymath}
induces the standard scalar product of quantum mechanics $\langle \psi|\tau\rangle =
\int dQ \,\psi^*(Q)\tau(Q)$ if the fixed 
state $\chi$ is normalized according to: $\int d\bar{Q}\,
\chi^*(\bar{Q})\chi(\bar{Q})=1$. 

Before going on let us notice that the request
of irreducibility is crucial in order to guarantee a one to one correspondence 
between the states of the theory and the vectors of the space ${\bf H}_{\chi}$
\cite{fonda}. In fact let us suppose we consider an arbitrary state belonging to ${\bf H}_{\chi}$
and we replace in it 
the function $\chi(\bar{Q})$ with a {\it different} square integrable function $\sigma(\bar{Q})$.
What we obtain is a {\it different} state of the KvN Hilbert space which corresponds
to the {\it same} expectation values for {\it all} the quantum observables
$\displaystyle f\biggl(Q,-i\hbar\frac{\partial}{\partial Q}\biggr)$, i.e. which
corresponds to the {\it same} physical situation. To avoid this 
redundancy we have to restrict the KvN Hilbert space to {\it only} one of the non-trivial 
subspaces which 
are invariant under the action of $\hat{Q}$ and $\hat{P}$. It is clear that the choice
of this invariant subspace is not unique. For example we could choose the
${\bf H}_{\chi}$ of (\ref{aro}) in which the wave function $\chi(\bar{Q})$ is fixed or the subspace 
\begin{displaymath}
{\bf H}_{\sigma}=\biggl\{ \psi(Q)\sigma(\bar{Q})\; \textrm{with} \, \int dQd\bar{Q}\,
|\psi(Q)|^2|\sigma(\bar{Q})|^2=1\biggr\}
\end{displaymath}
in which the wave function $\sigma(\bar{Q})$ is fixed
or every other subspace of the same form. Anyway it is easy to realize
that all these subspaces are isomorphic to the standard quantum Hilbert space of (\ref{quantumhil})
and that they are completely equivalent from a physical point of view. 


These considerations do not depend on the particular representation that we use. 
For example, in order to obtain 
the momentum representation of quantum mechanics, we have first of all to consider
the $(\lambda_q,p)$ representation of the KvN theory. In this representation 
$\hat{\lambda}_q$ and $\hat{p}$ are multiplicative operators, $\displaystyle 
\hat{q}=i\frac{\partial}{\partial \lambda_q},\hat{\lambda}_p=-i\frac{\partial}{\partial p}$ 
and the wave functions are of the form 
$\psi(\lambda_q,p)$ or, using the Bopp variables, $\psi(P,\bar{P})$.
On these wave functions the positions $\hat{Q}$ act as derivative operators:
$\displaystyle \hat{Q}=i\hbar\frac{\partial}{\partial P}$ while the 
momenta $\hat{P}$ act as operators of multiplication by $P$. Also in this 
representation if the operators
$\hat{Q}$ and $\hat{P}$ must act irreducibly on ${\bf H}$, then we have to identify 
${\bf H}$ with the space of all the square integrable functions which depend only on 
the quantum momenta $P$ times a particular $\chi(\bar{P})$. Such a space is again 
isomorphic to the usual quantum Hilbert space of the square integrable wave functions in the 
momentum representation $\psi(P)$.

Before concluding this section we want to stress 
that it is the ``quantization map" itself ${\cal Q}$ of Eq. (\ref{mapq}) 
which implies that we must change the operator of evolution by replacing 
the Liouvillian $\hat{L}$ with the operator $\hat{\cal G}$ in order to have a conserved quantum energy.
Moreover we must reduce the KvN Hilbert space to the usual Hilbert space of quantum mechanics 
in order that the positions $\hat{Q}$ and the momenta $\hat{P}$ constructed via
our map (\ref{mapq}) form an irreducible set of operators.
So the role of the map (\ref{mapq}) in this new approach to quantization 
of classical mechanics is crucial.

\section{Conclusions and Open Problems}

In this paper we have shown how to embed quantum mechanics in the KvN Hilbert 
space, which was first introduced to give an operatorial 
formulation of classical mechanics. In particular we have proved that the KvN Hilbert space
turns out to be a suitable mathematical framework for a unifying 
picture of classical and quantum mechanics: in fact among all 
the KvN Hermitian operators there are both the classical observables  
$f(\hat{q},\hat{p})$ and the quantum ones 
$\displaystyle f(\hat{Q},\hat{P})$, see the figure below.

\begin{figure}[!ht]
\centering
\label{kkk}
\includegraphics{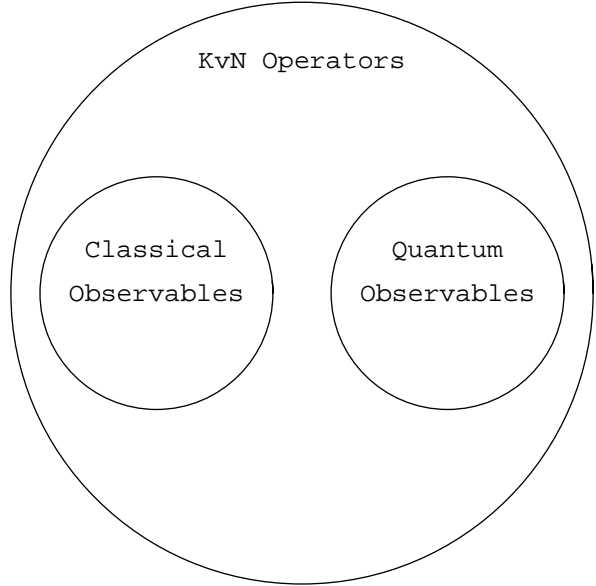}
\caption{\rm{\small{Classical and quantum observables can be seen 
as non-trivial subsets of all the possible Hermitian KvN operators.}}}
\end{figure}

\noindent As a consequence of the choice of the observables in classical mechanics only the 
moduli of the KvN wave functions $\psi(q,p)$ are physically 
significant \cite{kvn2} while in quantum mechanics, if we impose the condition that 
$\hat{Q}$ and $\hat{P}$ are an irreducible set of operators, the 
subspace of the physically significant KvN wave functions is isomorphic to the space 
of the square integrable wave functions
$\psi(Q)$ or $\psi(P)$, according to the representation we use.

Some ideas of the approach contained in this paper are similar to those 
of geometric quantization, see for example \cite{Woodhouse}. In both cases
the starting point is the KvN Hilbert space of square integrable functions 
on the phase space, in both cases 
a map is introduced to build the quantum observables and in both cases 
a sort of polarization is used to reduce the dimension of the quantum Hilbert space 
to the usual one. However in our approach 
the recipe to construct the observables is very simple and far from
the complicated tools used in geometric quantization. The simple map 
(\ref{mapq}) itself suggests in fact the two other ingredients needed to
quantize a classical system. The first one is a well-motivated 
physical prescription, the conservation of the quantum energy, 
which implies that the dynamics of the theory has to be modified.
The second ingredient is the mathematical prescription which says that the operators
$\hat{Q}$ and $\hat{P}$ must act irreducibly and which is necessary to 
eliminate the redundancy in the physical description of the system.
Anyway, like in every other approach to quantization also in our case we cannot bypass 
the Gr\"oenewald-van Hove theorem \cite{Gro} and an obstruction to a full quantization 
\cite{Abraham} of the system 
still remains. In geometric quantization people construct a very complicated 
$\wedge$-map just to satisfy the relation 
\begin{equation}
\label{kit}
\displaystyle \{f,g\}_{Pb}\;\hat{\longrightarrow}\;\frac{1}{i\hbar} [\hat{F},\hat{G}]
\end{equation}
for every function $f$ and $g$
but, when they impose the request of irreducibility, the number of observables which can 
be quantized gets limited. In our approach
the irreducibility prescription reduces the KvN Hilbert 
space to the usual quantum one in a more natural way than in geometric
quantization and again this is motivated by the a priori choice of the observables
(\ref{mapq}) that we made. Nevertheless our map (\ref{mapq}) cannot satisfy
Eq. (\ref{kit}) for every function 
$f$ and $g$ as a simple calculation can show. Therefore the 
Gr\"oenewald-van Hove theorem is not bypassed by our quantization method. 

The last point we want to mention is that having a unique 
big Hilbert space for both classical and quantum mechanics can help in studying and appreciating 
the similarities and differences between the two theories. In particular, for what concerns
the observables 
we must note, see {\bf Fig. 1}, that within the KvN theory there is enough space 
to describe both classical and quantum mechanics but there is further space available 
for further observables which could describe some other possible regimes, like
the ones at the border between classical and quantum mechanics \cite{physreva}
or the models of dynamical reduction \cite{grw}. We think that it would be interesting to investigate 
these issues in the future.

\section*{Acknowledgments}
I would like to thank E. Gozzi for his precious help and suggestions.
This research has been supported by grants from INFN, MIUR and the University of Trieste.


\begin{thebibliography}{99}
\bibitem{Koopman}
B. O. Koopman, Proc. Natl. Acad. Sci. U.S.A. 17 (1931) 315.
\bibitem{von Neumann}
J. von Neumann, Ann. Math. 33 (1932) 587; ibid. 33 (1932) 789.
\bibitem{gozzi}
E. Gozzi, M. Reuter, W. D. Thacker, Phys. Rev. D 40 (1989) 3363.
\bibitem{waves1}
D. Mauro, Int. J. Mod. Phys. A 17 (2002) 1301.
\bibitem{quantoptics}
J. E. Moyal, Proc. Camb. Philos. Soc. 45 (1949) 99.\\
For a review of the Wigner function formalism, see:\\
W. P. Schleich, Quantum Optics in Phase Space, Wiley-VCH, Berlin, 2001.
\bibitem{kvn2}
E. Gozzi, D. Mauro, to appear.
\bibitem{marinov}
M. S. Marinov, Phys. Lett. A 153 (1991) 5.
\bibitem{ennio}
E. Gozzi, M. Reuter, Int. Jour. Mod. Phys. A 9 (1994) 2191.
\bibitem{Kubo}
R. Kubo, Jour. Phys. Soc. Jap. 19 (1964) 2127.
\bibitem{ann}
E. Gozzi, D. Mauro, Ann. Phys. 296 (2002) 152.
\bibitem{Abraham}
R. Abraham, J. E. Marsden,  Foundations of Mechanics, Benjamin, New York, 1978.
\bibitem{Woodhouse}
N. M. J. Woodhouse, Geometric Quantization, Oxford University Press, Oxford,
1980.
\bibitem{fonda}
L. Fonda, G. C. Ghirardi, Symmetry Principles in Quantum Physics, Dekker, New York, 1970. 
\bibitem{Gro} 
H. J. Gr\"oenewald, Physica 12 (1946) 405\\
L. van Hove, Mem. de l'Acad. Roy. de Belgique (Classe des Sci.) 26 (1951) 61.
\bibitem{physreva}
L. Djosi, N. Gisin, W. T. Strunz, Phys Rev. A 61 (2000) 022108\\
O. V. Prezhdo, V. V. Kisil, Phys. Rev A 56 (1997) 162\\
A. Peres, D. Terno, Phys Rev. A (2001) 022101\\
J. Caro, L. L. Salcedo, Phys. Rev. A (1999) 842.
\bibitem{grw}
G. C. Ghirardi, A. Rimini, T. Weber, Phys. Rev. D 34 (1986), 470.
\end{thebibliography}
\end{document}